# Generalized Achlioptas process for the delay of criticality in the percolation process.

Paraskevas Giazitzidis and Panos Argyrakis

Department of Physics, University of Thessaloniki, GR-54124 Thessaloniki, Greece

**Abstract**

In this work we extend the Achlioptas model for the delay of criticality in the percolation problem. Instead of having a completely random connectivity pattern, we generalize the idea of the 2-site probe in the Achlioptas model for connecting smaller clusters, by introducing two new models: The first one by allowing any number k of probe sites to be investigated, k being a parameter, and the second one independent of any specific number of probe sites, but with a probabilistic character which depends on the size of the resulting clusters. We find numerically the complete spectrum of critical points and our results indicate that the value of the critical point behaves linearly with k after the value of k = 3. The range k = 2-3 is not linear but parabolic. The more general model of generating clusters with probability inversely proportional to the size of the resulting cluster produces a critical point which is equivalent to the value of k being in the range k = 5-7.

**Introduction**

The percolation phase transition has attracted considerable interest over the years because it can be considered as a fundamental model of a nonequilibrium system [1]. It is well known that it has several practical applications in many scientific fields. The main characteristic of this system is the sudden appearance of the largest percolating cluster at the critical point, causing the system to go suddenly from one state (insulator) to the other (conductor). Very recently the problem of generating percolation clusters at the critical point has attracted renewed attention, especially after the finding of the Achlioptas process [2], which is a technical method for producing a delay in the appearance of the infinite percolation cluster. The problem can be realized either in networks [2] or in lattices [3-6]. In the present work we examine it in 2D and 3D lattices. The normal percolation problem is attained by starting with a lattice with all empty sites (bonds), and then filling each site (bond) randomly one at a time, paying attention to the generation of clusters, especially at the appearance for the first time of the single infinite percolating cluster, which spans the lattice (or network) from one end to the other. In the Achlioptas process one does not fill single sites (bonds), one at a time, as the usual percolation problem described above, but instead, it initially fills randomly two sites (bonds), then investigates the sizes of the two new joined clusters (either sum or product of the sizes of the constituent joining clusters), and finally it keeps only the site (bond) that results in a smaller cluster and discards the other one. This procedure has as a result the formation of smaller clusters, as compared to the normal case of adding one site (bond) at a time, and as a consequence the delay of the appearance of the critical point. For the case of site percolation in two dimensions the Achlioptas critical point is 0.695 instead of the usual 0.593 value [4], while it is 0.525 instead of the usual 0.500 for the case of bond percolation [5-6].

Equivalent results can be obtained for different lattice geometries.

Here, we extend the Achlioptas process and instead of adding two new sites (bonds) at a random location, we now add k such sites (bonds), k being an integer, e.g. k=3,4,5… We then follow the Achlioptas process and choose only one of the k sites (bonds), the one that results to the production of the smallest cluster, while all others are removed. We use the same idea of the delay of criticality and we evaluate numerically the critical point for different values of k. As expected, we find that the larger the k value, the greater the delay for criticality. The motive for investigating this problem is to find how is the delay to criticality affected by changing the value of k, i.e. to see if such extended Achlioptas process behaves linearly or non-linearly in the entire k range. Of course, being able to manipulate the exact location of the critical point is interesting in itself when designing systems or processes that need to control the critical point.

Furthermore, by generalizing this idea we introduce a new model for the cluster generation, which does not use at all the idea of probing a number of k possibilities, but increase the value of p by use of a probabilistic model, which is the following: Since the main idea is to delay the emergence of the largest cluster by retaining only small clusters we now increase p by adding a single site (bond) at a random location and by checking the clusters that this site (bond) happens to attach to, or the two clusters that it now connects. If the size of the resulting cluster is S, then we will retain the newly added site (bond) with probability 1/S. We expect the same type of behavior as the original Achlioptas procedure, but now in a generalized manner, independent of k.

Of considerable interest in the past two years has also been the characterization of the new process as a phase transition of the first order, something that recently drew considerable attention, with ambivalent opinions been expressed in the literature [6-12], regarding the character of this new transition. While the original assumption [2] has been that indeed the Achlioptas process leads to a first order transition, as opposed to the classical percolation which is a second order process, there is now evidence that this is not really true [8]. Very refined numerical calculations of extremely large size systems are needed to prove or disprove the original assumption. The findings in this work are not affected by this characterization.

**Method of calculations**

First we calculate the bond percolation characteristics of a 2D square lattice by using different ways for the construction of the clusters. Initially, we have an empty lattice of size $L^2$. In the first model, we start by placing at random in each step a number (k) of independent possible bonds which are connecting the nodes on which they are placed according to the following rules. The number of bonds k that we place on the lattice is a parameter and a key element of this study. For example, in Achlioptas process the parameter k=2. Initially, as no clusters exist, we randomly choose one of these k bonds, and we allow it to remain on the lattice, while we remove the rest. As the concentration of bonds is increased the newly added bonds may start to connect small clusters to form a larger one. When this happens, we calculate for each possible bond the product of the sizes of these small clusters that are going to be merged into larger ones. We do this for all k possible bonds added in one step. Then, we retain only one bond, the one that results in the smallest product of the sizes of the clusters to be merged. We then, take out the other (k-1) bonds that resulted in larger products. We continue the process until the lattice is full.

   We then investigate the Achlioptas process in a simple cubic lattice. We initially estimate the value of the critical point by monitoring the size of the largest cluster as we change the probability density of the bonds in the lattice. Finally, we do the same investigation for the 2D triangular lattice and we observe the same behavior for the delay of the critical point as in the 2D square lattice.

In a second model that we use, we start with an empty lattice of size $L^2$ by placing at random a single bond on the lattice and we investigate the size of the clusters that this bond would be connected to. Let us say, that these two clusters have size $S_1$ and $S_2$. We calculate $S = S_1 + S_2$. Then we decide to either retain or remove this new bond. We retain it with probability $1/S$. Note that in the case the newly added bond does not connect to a cluster on any end, then the contribution to the sum S is simply 1 for that end. We also continue the process until the lattice is filled up.

**Results**

We calculate the quantities $I_{av}$ and $I'_{av}$ as a function of p. $I_{av}$ is the average size of all clusters in the lattice

$$I_{av} = \sum_{m=1}^{m_{max}} \left(\frac{i_m \cdot m^2}{p \cdot N^2}\right) \qquad (1)$$

where $m$ is the size of the cluster, $i_m$ is the number of clusters of size $m$, $p$ is the concentration of bonds and $N$ is the maximum number of bonds.

$I'_{av}$ is the average size of all clusters minus the size of the largest cluster.

$$I'_{av} = \sum_{m=1}^{m-m_{max}} \left(\frac{i_m \cdot m^2}{p \cdot N^2}\right) \qquad (2)$$

These two quantities are characteristic of the percolation phase transition, clearly showing the sharp change. $I_{av}$ has a fast rise around the critical point, while $I'_{av}$ has a sharp peak at the critical point, and this is very useful to identify it. We do this for several different values of k, in the range $k = 2 - 20$. In Figure 1 we show the curves for $I_{av}$ and in Figure 2 the curves of $I'_{av}$ for the 2D square lattice. We can now identify the location of the critical point, which is in the middle of the rise of the curve in Figure 1, or the peak in Figure 2. As expected, the curves move to the right as k increases. This produces an increasing delay of $p_c$ from 0.526 in the case of k=2 to 0.592 in the case of k=20.

In Figure 3 we plot $I'_{av}$ as a function of the probability density of bonds for a 3D simple cubic lattice of size $L^3=200^3$. The graph shows that the critical density by probing two independent random bonds (k=2) is $p_c=0.32$. This value of $p_c$ is not in good agreement with the work of S. Fortunato et al [5], who report a value of $p_c=0.3876$.

For this reason we use an additional method based on Finite Size Scaling to calculate the critical point for the case of k=2. In this method we investigate the scaling of the size of the giant percolating cluster ($S_{max}$) in the region near the critical point, for several different values of the lattice size $L^3$. This dependence is shown in Equation 3.

$$S_{max}(L) = P_{max} \cdot L^3 = L^{d_f} \qquad (3)$$

where $P_{max}$ is the size of the giant component normalized by the system size, and it gives the probability for a random site in the lattice to belong to the giant percolating cluster. In Equation 3 $d_f$ is the fractal dimension, which shows how the size of the giant component $S_{max}$ scales with L.

In Figure 4 we plot $S_{max}$ as a function of L, for four different bond concentrations. The value of $p_c$ will be given by the curve that is a straight line, so that $S_{max}$ depends on L by a power law. From Figure 4 we conclude that $p_c = 0.322$. From the slope of the power law we extrapolate the value of the exponent $d_f = 2.76 \pm 0.02$. In Figure 5 we plot the mean cluster size S near $p_c$ as a function of L for five different bond concentrations. This dependence is shown in Equation 4:

$$S(L) = L^{\gamma/\nu} F[L^{1/\nu}(p - p_c)] \qquad (4)$$

where $\gamma$ and $\nu$ are the universal critical exponents. From Equation 4, when $p=p_c$, $S(L)$ scales logarithmically with L. As we observe from Figure 5 the straight line in the log-log diagram appears at the density p=0.322 while all the other nearest values of p do not have a logarithmic law scaling with L. From Figure 5, we conclude that the value of the critical density of bonds is $p_c = 0.322$ and the critical exponent $\gamma/\nu = 2.487 \pm 0.01$ as the slope of the straight line. In Figure 6 we show the dependence of the variance $\sigma^2(p)$ of the size of the giant cluster near $p_c$ on the density p for different lattice sizes. From the peaks of the curves in Figure 6 we conclude that the critical density of the product rule in a simple cubic lattice is $p_c=0.322 \pm 0.0005$, in good agreement with our previous method.

In Figure 7 we plot $p_c$ as a function of k for the range k=2 to k=20 for all three different types of lattices we have investigate (square, triangular, simple cubic). We observe here a linear relation for the range k= 3 – 20. However, this is not the case for the first point, k=2, which gives a considerably lower $p_c$ value, meaning that the effect of the delay of criticality is not so pronounced as for the cases of k ≥ 3. We investigate in detail the range k = 2 – 3 by considering several points in this special interval. We do this by taking fractional values of k. This is done by changing k continuously during the cluster generation between the values of interest. For example, for the case of k=2.5 we consider at one instance two candidate bonds, the next step three bonds, the next one again two bonds, etc so that on the average a number of k=2.5 bonds is been considered. Similarly for the other fractional values. As we can observe in Figure 7, at the square and triangular lattices $p_c$ is increasing proportionally with k, with slope of 0.003, instead of the simple cubic lattice where the proportional increase has a slope of 0.0006.

In Figure 8 we show $I_{av}$ and $I'_{av}$ for the case of the second model where only one bond is added with probability inversely proportional to the size of the cluster. The data given pertain to a 2D square lattice. This model gives a value of $p_c=0.545$. Curiously enough, the critical point for this model is in the range k=5-7 of the first model.

**Conclusions and Summary**

We have investigated in detail the percolation problem by employing a generalized Achlioptas process and evaluating numerically the values of the critical point in the phase transition. This process involves using a variable number of probe sites, as opposed to the original work which used two sites only.

There were three different types of lattices that have been studied in this work (square, triangular, simple cubic). Both 2D lattices for k=2 (Achlioptas process) are in a very good agreement with previous publications. For the 3D simple cubic lattice there still remains some disagreement for the position of the critical point for the classical Achlioptas processes. However, the gist of this work is Figure 7 which shows the position of the critical point as a function of k. Our results clearly show the delay of criticality as k increases. A second model based on the probability of growth of the clusters was also used, with the expected delay of the critical point. This model gave equivalent results to the first model.

Another similar work of probing one bond according to a probability rule was reported by Schrenk et al [13]. In this model a different criterion was used, which was based on the difference between the size of the cluster that this bond merges with, and the mean cluster size of the system.

Another so called generalized Achlioptas process has been recently introduced by M. X. Liu et al and J. Fan et al [14-15]. In this method, two unoccupied bonds are chosen randomly in the system. The one that minimizes the product of the sizes of the clusters that will be merged is been retained, but with a probability p (not to be confused with our "p" parameter, which is the usual probability for an occupied site). At p = ½, this model is equivalent to the classical ER network where bonds are picked up randomly [1]. At p = 1 the rule becomes equivalent to Achlioptas process [2]. The method described above has been implemented in both random networks and 2D lattices [14-15], which resulted that this phase transition, independently of p parameter, is a continuous one. These two works also report the dependence of the critical point on this new parameter p.

Such approaches that produce a succession of several critical points depending on the method of preparation of the percolation system are useful in that they give a control over the critical phenomena. One can vary the point of the transition by suitably using different numbers of probe sites. For this reason there are several other strategies reported in the literature leading to different modifications of the percolation threshold. A comprehensive list of over 120 papers devoted to percolation thresholds estimation is available at: http://en.wikipedia.org/wiki/Percolation_threshold [16].

Acknowledgment: This work has been supported the European Commission projects INTERREG ICoSCIS-B2.33.02, and FP7 project MULTIPLEX 317532. We thank Dr. J. Hoshen for several useful discussions and for the suggestion to investigate the second model in this work. We also thank Prof. S. Fortunato for discussions.

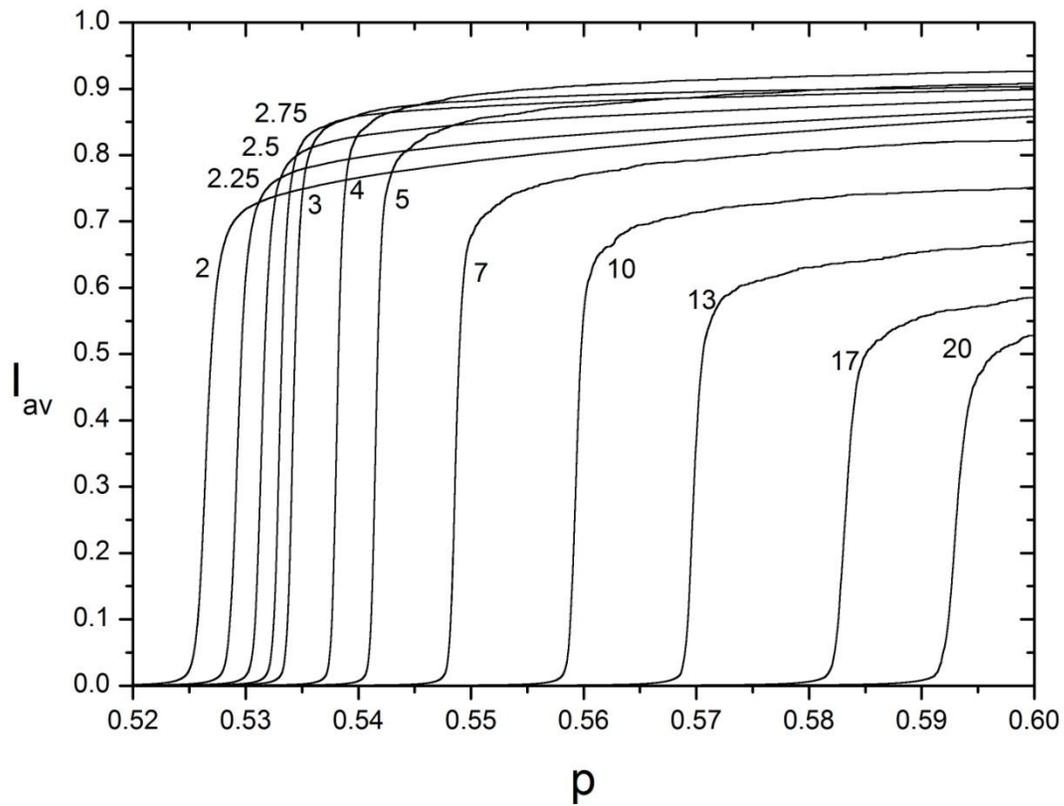

Figure 1: Plot of the average cluster size $I_{av}$ as a function of the concentration of bonds p, for different k values. The results have been averaged over $10^3$ realizations for a 2D square lattice of size $L^2 = 1000^2$.

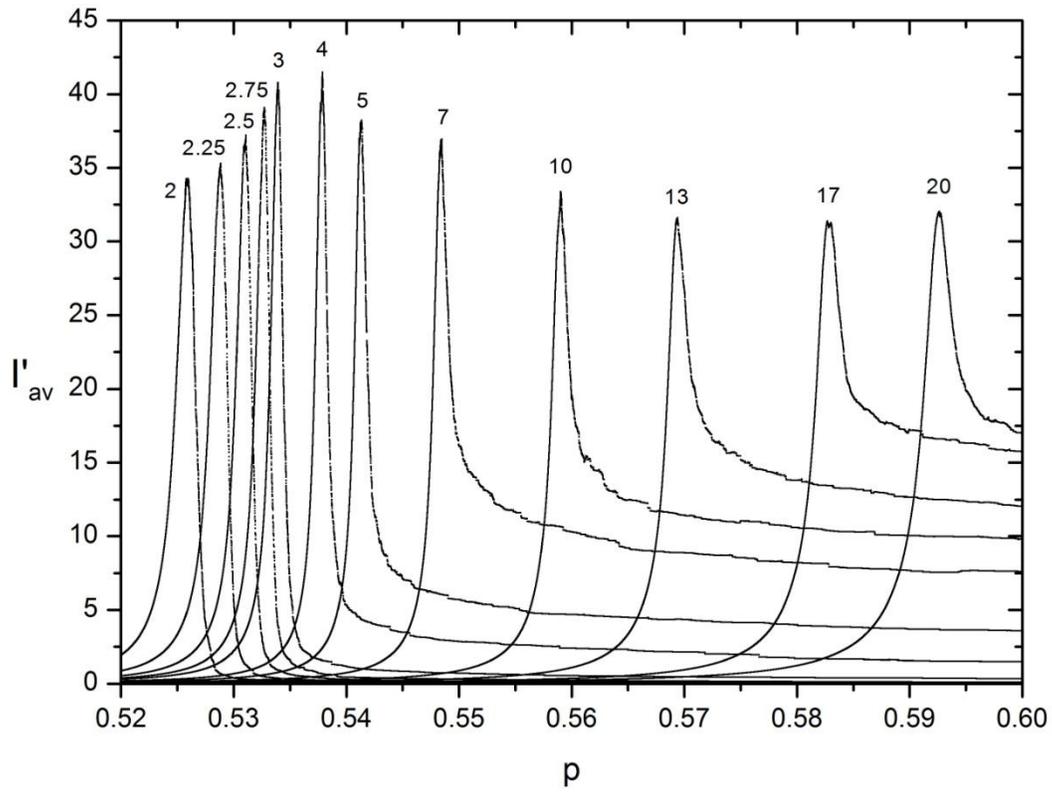

Figure 2: Plot of the average cluster size minus the largest cluster I'$_{av}$ as a function of the concentration of bonds p, for different k values. The results have been averaged over $10^3$ realizations for a 2D square lattice of size $L^2 = 1000^2$.

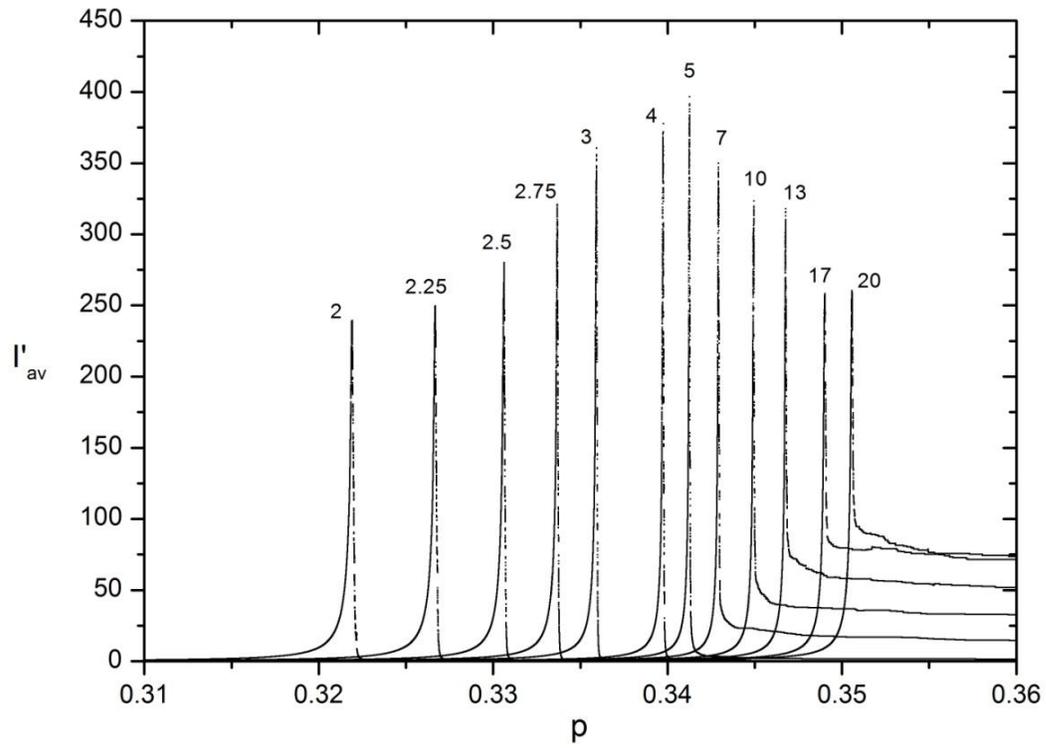

Figure 3. Plot of I'$_{av}$ versus the concentration of bonds p for a simple cubic lattice of size L$^3$ = 200$^3$, and for different values of k.

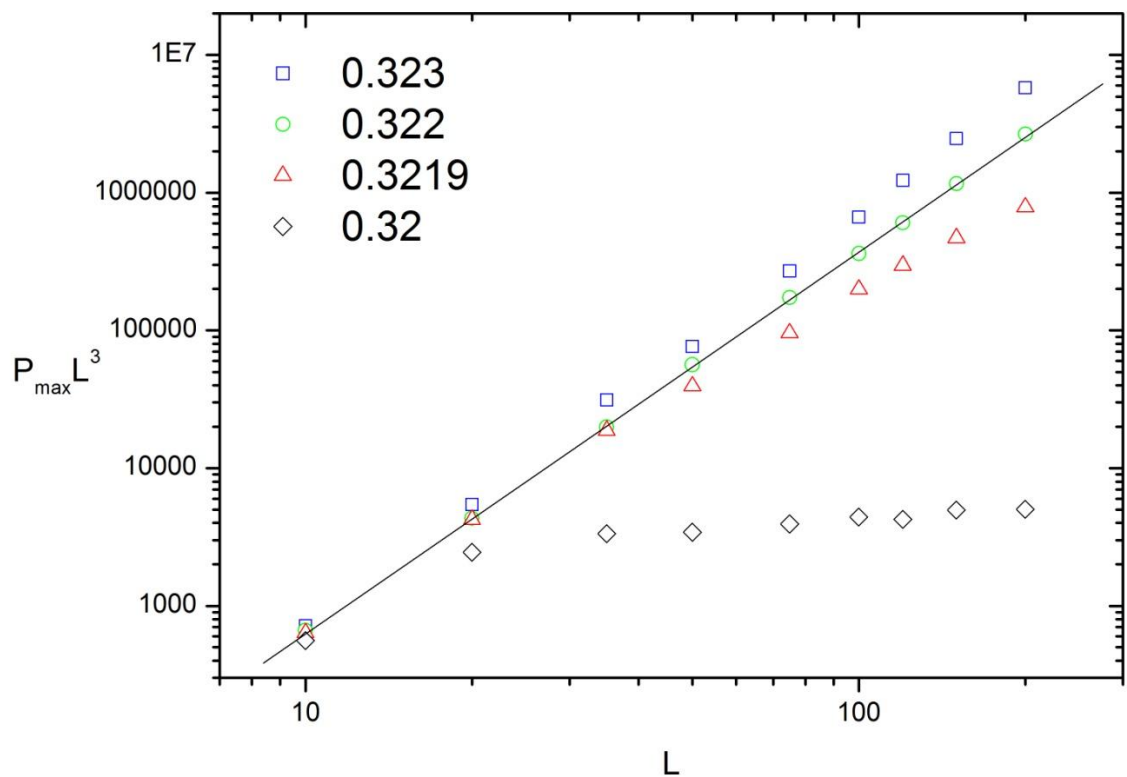

Figure 4: Finite size scaling plot of $P_{max}L^3$ versus L for a simple cubic lattice, for four different bond concentrations.

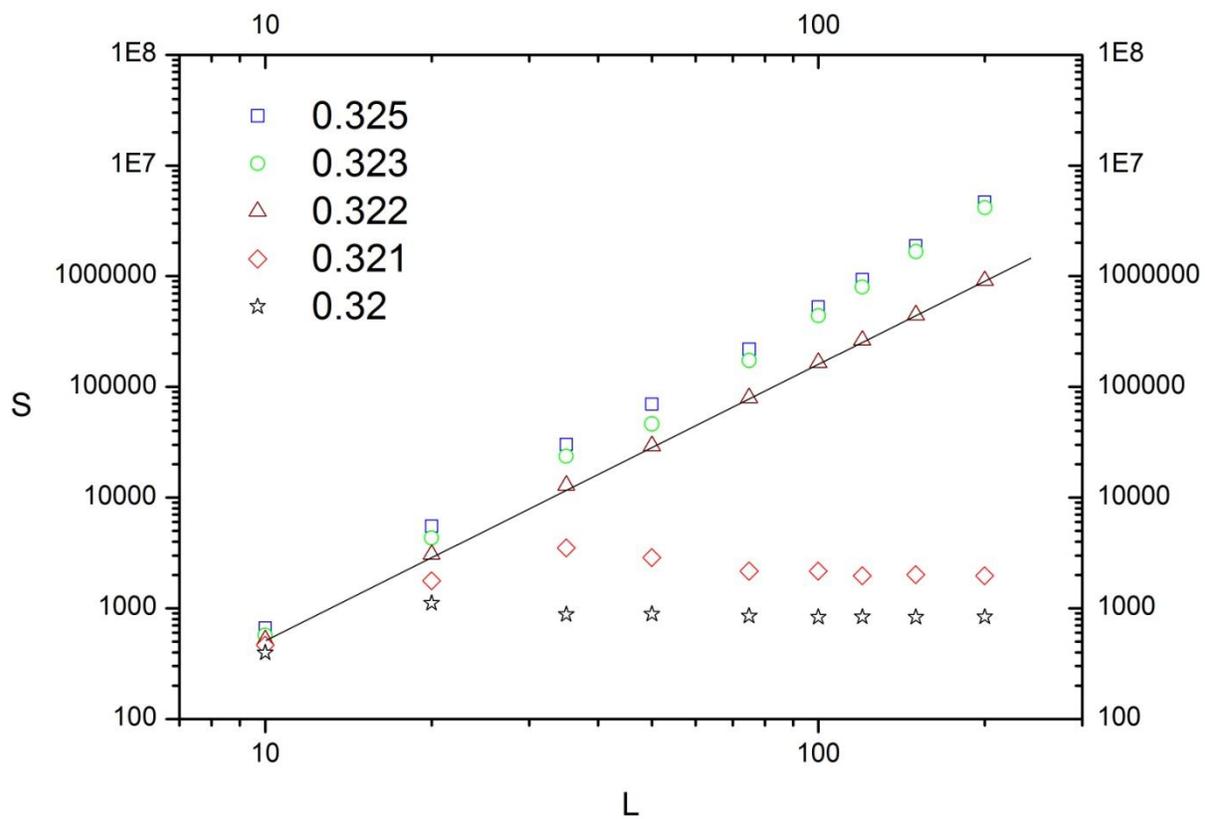

Figure 5. Finite size scaling plot of S versus L for a simple cubic lattice, for five different bond concentrations.

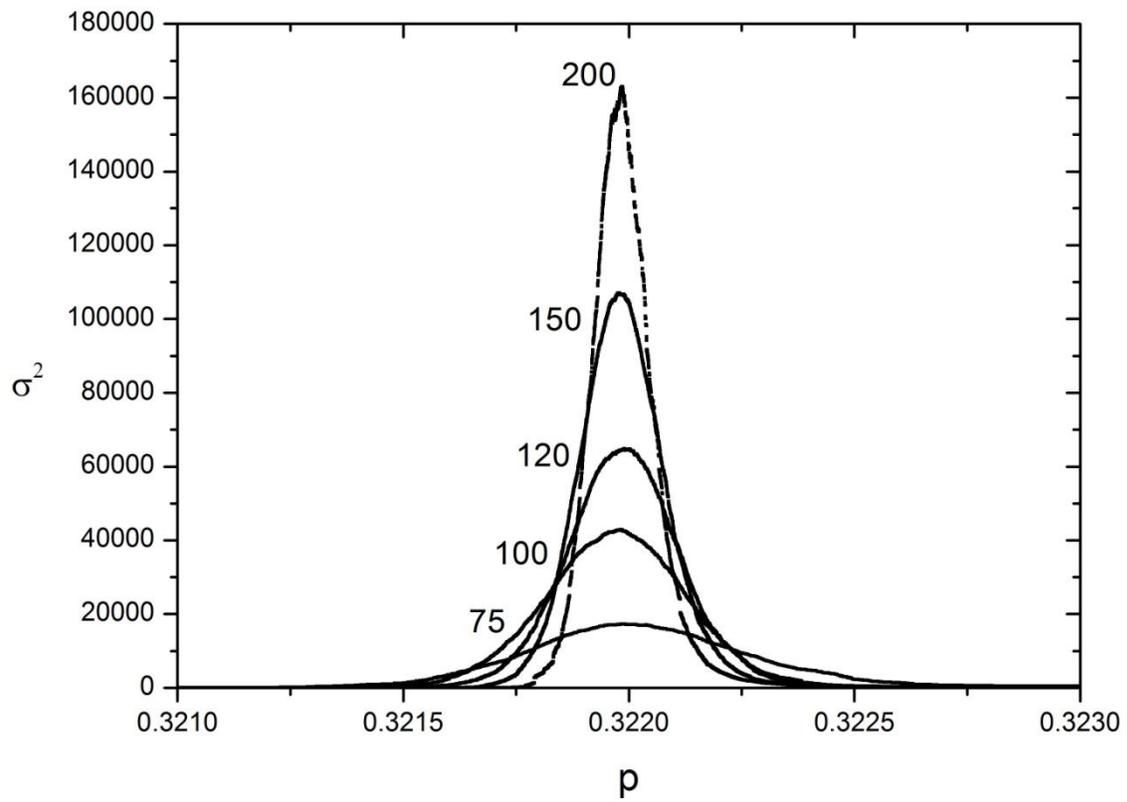

Figure 6. Plot of the variance of the largest cluster versus bond concentration p for seven different simple cubic lattices with sizes L$^3$ from L = 75(lower pick) to L = 200(higher pick).

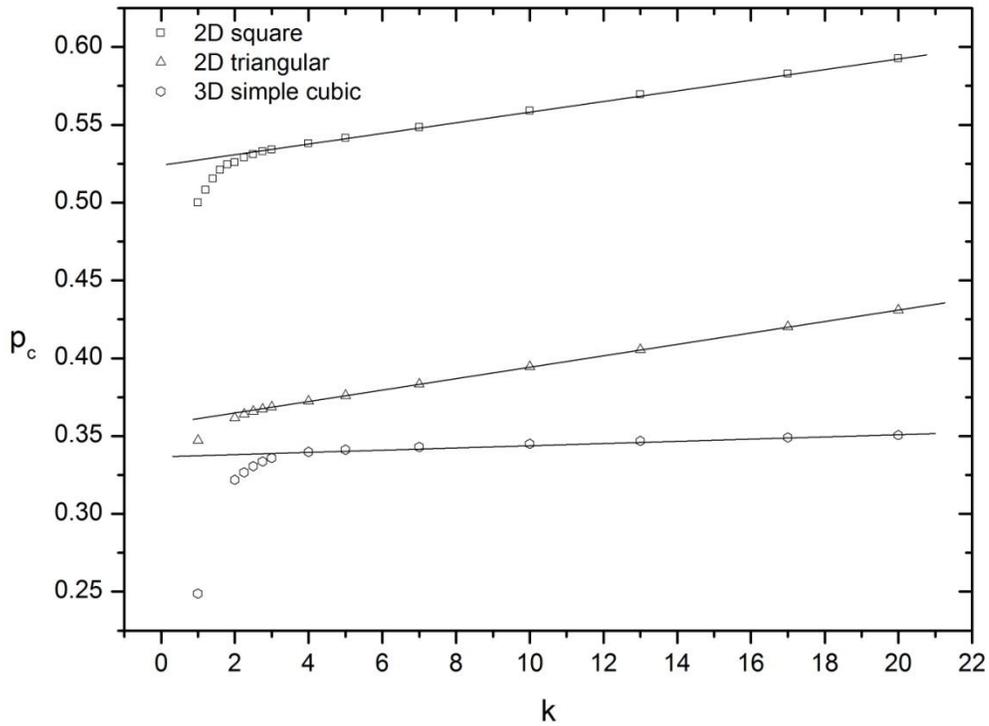

Figure 7: Plot of $p_c$ as a function of k for three different lattices. The straight solid line is one of slope 0.003 for the square and the triangular lattices and 0.0006 for the simple cubic lattice. The lattice size is $L^2=1000^2$ for the 2D lattices and $L^3=200^3$ for the simple cubic, while all data have been averaged over $10^3$ realizations. The first five points of the square lattice curvature have been taken from Reference 14.

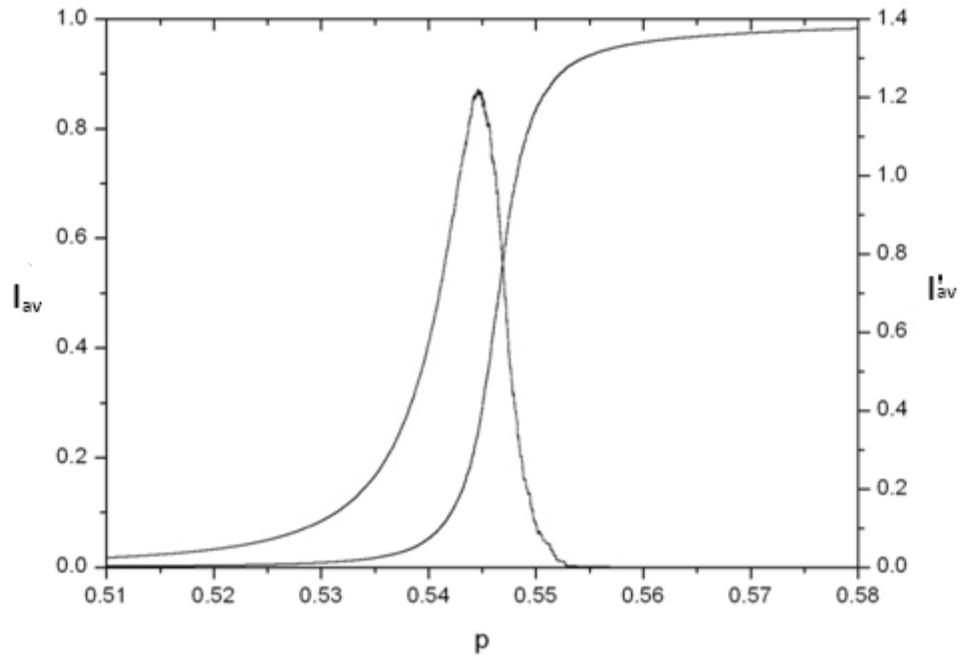

Figure 8: Plot of the $I_{av}$ (left axis), and $I'_{av}$ (right axis) of a bond percolation system in which the clusters are merging with a probability of $\frac{1}{S}$ (where S is the size of the resulting cluster after the merging). The results have been averaged over $10^3$ realizations for a system size of $L^2 = 300^2$ (see text for details).